\documentclass[aps,pra,reprint,amsmath,amssymb,floatfix,longbibliography,raggedbottom]{revtex4-2}
\usepackage{mathtools,booktabs,graphicx,microtype,xcolor,hyperref,xurl,needspace}
\hypersetup{colorlinks=true,linkcolor=blue!40!black,citecolor=blue!40!black,urlcolor=blue!40!black,pdfauthor={Jaebum Kim}}
\graphicspath{{figures/}}
\newcommand{\fig}[3]{\begin{figure*}[t]\centering\includegraphics[width=\textwidth]{#1}\caption{#2}\label{#3}\end{figure*}}
\newcommand{\referencefont}{\footnotesize\linespread{0.98}\selectfont}
\newcommand{\referencespacing}{\setlength{\itemsep}{1pt}\setlength{\parsep}{0pt}}

\begin{document}
\title{Circuit-aware protection against burst errors in lattice surgery}
\author{Jaebum Kim}\noaffiliation
\begin{abstract}Large-scale quantum computing will require logical operations protected against high-energy impacts that disturb many superconducting qubits at once. Gap engineering suppresses relaxation induced by these impacts, but protecting only a subset of qubits requires choosing where that protection is placed. During lattice surgery, measurements change as patches merge and split, exposing different ancillas and temporal boundaries to burst errors. Here we compare three static protection layouts for a simulated lattice-surgery CNOT, matching the numbers of protected ancillas and the X:Z check-type splits. At distance 9, protecting the check type measured during each merge reduces burst-induced excess risk by $38^{+6}_{-5}\%$ (95\% bootstrap interval) relative to a deep geometric band under correlated background matching, without giving the decoder the impact parameters. We attribute this difference to report errors at temporal boundaries and ancilla faults propagating to data, which interact to increase the excess risk of a shallow band. We also find that the typed layout reaches the shallow-band reference risk with an estimated $49^{+2}_{-14}$ fewer protected ancillas. At distance 13, the advantage is not resolved under oracle decoding, and a five-round burst cap reverses the shallow-band ordering. These results indicate how the measurements of a logical circuit can guide protection placement within the simulated decoder and burst-duration regimes tested here.\end{abstract}
\maketitle
\widowpenalty=10000\clubpenalty=10000
High-energy impacts in superconducting-qubit arrays cause correlated relaxation and charge errors at multiple sites~\cite{mcewen,wilen,martinis}. Gap engineering suppresses the relaxation induced by an impact~\cite{gap}, although correlated phase-error bursts persist in gap-engineered arrays~\cite{phase}. An impact can also corrupt an ancilla measurement or induce a fault that propagates from an ancilla to data qubits, changing the syndrome used for correction. Previous approaches move logical information away from an impact~\cite{flight}, remap magic-state factories~\cite{factories}, adapt the geometry of a code to defects~\cite{deformer} or exploit the tolerance of surface codes to short bursts~\cite{burstresilience}. These approaches leave open where to place static protection as the measured checks change during lattice surgery.

When surface-code patches merge or split, the checks being measured change and create temporal boundaries in the syndrome record~\cite{surgery,temporalsurgery,fragile}. These boundaries can make an ancilla near a merge seam more important than another ancilla at the same distance from the seam. Here, we compare three static protection layouts for a lattice-surgery CNOT using simulated impacts and four decoders, holding protected-ancilla counts and X:Z check-type splits constant across layouts while examining how measurement faults and ancilla-to-data propagation affect logical risk as the circuit proceeds. At distance 9, the layout protecting the measured check type has 38\% lower burst-induced excess risk than a deep geometric band under correlated background matching. We then test how many protected ancillas are needed to reach a common reference risk and how the ordering changes with decoder information and burst duration.

\section{Lattice-surgery CNOT and protection layouts}
We simulate a lattice-surgery CNOT on three surface-code patches arranged in an L shape, with control C, helper A and target T, and measure joint logical parity in consecutive $ZZ$ and $XX$ merges~\cite{surgery}, followed by a $Z$ readout of A. For each impact, we compute the dose at every site from its energy, position and onset, map those doses to Pauli, report and reset errors in the presence of background circuit noise~\cite{si1000}, and sample the resulting faults in Stim~\cite{stim}. We test two complementary input states and count a logical failure when any tested output is flipped (Methods).

We keep each protection layout static throughout the CNOT and compare three ways of protecting ancillas at matched budgets (Fig.~\ref{fig:layouts}). Specifically, the typed layout protects Z-check ancillas near the $ZZ$ seam and X-check ancillas near the $XX$ seam, counting depth in layers of the measured check type, whereas the shallow and deep geometric bands protect both check types according to distance from the seam, with the deep band extending as far as the budget permits. In every layout, data qubits and sites used only during a merge are strongly protected, and the counts of protected X and Z checks are matched separately. At distance 13, both the deep and typed layouts protect all measured-type checks at the seam, although their weak ancillas occupy different regions (Fig.~\ref{fig:layouts}).

\begin{table*}[t]
\caption{Protection layouts and matching decoders. Correlations refer to joint detection events. The oracle uses realized simulated impact parameters without access to the fault history.}\label{tab:definitions}
\begin{tabular}{@{}p{0.25\textwidth}p{0.70\textwidth}@{}}\toprule
Term & Definition\\\midrule
Typed band & Protects the measured check type near each merge seam, with width counted in layers of same-type checks.\\
Shallow geometric band & Protects both check types within a constant shallow depth of the seams.\\
Deep geometric band & Protects both check types within the largest possible depth at the same protected count.\\
Memory boundary layout & Leaves weak Z checks nearest the horizontal X boundaries of each patch and weak X checks nearest its vertical Z boundaries, with matched counts and a deterministic tie order.\\
Background matching, $D_{\rm bg}$ & Uses the burst-free noise model.\\
Background + correlations, $D_{\rm bg}^{c}$ & Uses the burst-free model and joint detection events.\\
Event-informed oracle, $D_{\rm or}$ & Adjusts the matching weights using the realized impact.\\
Oracle + correlations, $D_{\rm or}^{c}$ & Uses realized-impact weights and joint detection events.\\\bottomrule
\end{tabular}
\end{table*}

\fig{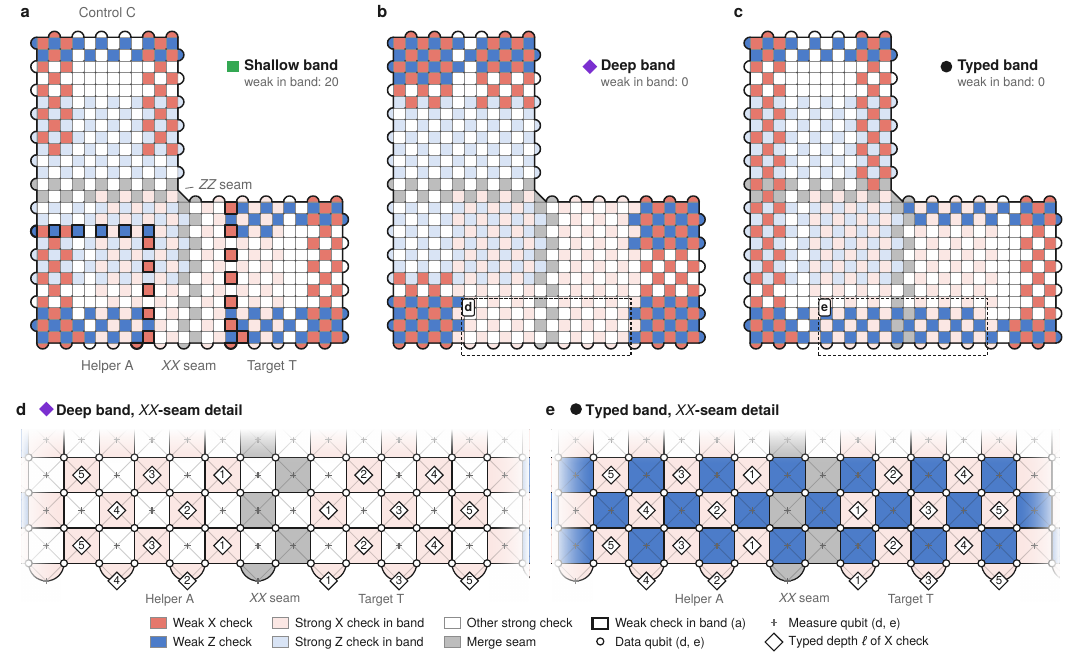}{\textbf{Protection layouts at a matched budget.} \textbf{a}--\textbf{c}, Shallow (\textbf{a}), deep (\textbf{b}) and typed (\textbf{c}) layouts at distance 13. Each has 153 weak ancillas among 532 eligible measurement ancillas. Weak X checks are red and weak Z checks blue; strong measured-type checks within seven typed seam layers are light red or blue. Other strong checks are white, and merge-only checks are grey. Outlines mark weak checks inside the measured-type band: 20, 0 and 0 in \textbf{a}--\textbf{c}, respectively. Small dots represent data qubits. \textbf{d},\textbf{e}, Enlarged $XX$ seam in the deep (\textbf{d}) and typed (\textbf{e}) layouts. Deep protects both check types; typed protects measured X checks with adjacent Z checks weak. Numbered diamonds mark typed depth $\ell$. Data qubits remain strongly protected.}{fig:layouts}

We decode each simulated measurement record with four configurations of PyMatching~\cite{pymatching} (Table~\ref{tab:definitions}): background or event-informed weights, the latter conditioned on the realized impact but not the sampled fault history, each evaluated with and without correlations between joint detection events~\cite{correlated}. These decoders let us compare placement when the decoder knows only background noise or is provided with the impact. For the two input modes, we define excess risk as the larger increase in logical failure probability above the quiet baseline, conditional on an impact (Methods).

\section{Placement comparisons and protection counts}
We first compare the typed and deep layouts in a distance-9 CNOT over 3,200 simulated impacts and find lower excess risk for typed under all four decoders (Fig.~\ref{fig:placement}a). For the ratio of deep to typed excess risk, denoted by $\eta_D$, correlated background matching gives $\eta_{D_{\rm bg}^{c}}=1.62^{+0.15}_{-0.13}$. This corresponds to a $38^{+6}_{-5}\%$ reduction in excess risk when the decoder receives no information about the impact. Here, the uncertainties represent 95\% bootstrap confidence intervals (Methods). Notably, with an event-informed oracle and correlated matching, $\eta_{D_{\rm or}^{c}}=1.43$, giving 30\% lower risk for typed and indicating that decoder mismatch accounts for only part of the placement effect. The other two decoders give ratios of 1.36 and 1.28 (Fig.~\ref{fig:placement}a), with the same ordering.

\begin{table}[t]
\caption{Protection-count sweep at distance 9 using correlated background matching. Excess risk is conditional on an impact. Intervals show nominal 95\% paired event-bootstrap estimates including quiet-baseline uncertainty. Counts below 117 are not possible for the five-layer typed band at the selected type budgets.}\label{tab:count-refinement}
\begin{ruledtabular}
\begin{tabular}{lcrl}
Layout & Protected sites & \multicolumn{2}{c}{Excess risk ($10^{-4}$)}\\
 & & Estimate & 95\% interval\\\hline
Typed & 117 & 8.29 & 6.91--11.64\\
Typed & 123 & 6.19 & 5.22--7.15\\
Typed & 132 & 5.10 & 4.22--5.98\\
Typed & 150 & 3.07 & 2.39--3.71\\
Typed & 168 & 2.42 & 1.74--3.08\\
Typed & 204 & 1.11 & 0.54--1.64\\
Shallow & 168 & 7.58 & 5.00--10.55\\
\end{tabular}
\end{ruledtabular}
\end{table}

To determine whether the typed layout reaches the same logical risk with fewer protected ancillas, we use a shallow band protecting 168 of 240 eligible ancillas (70\%) as the reference in a model that also permits full protection (Fig.~\ref{fig:placement}b). We compare its risk against typed layouts with different numbers of protected ancillas, including a five-layer typed band at the lowest permitted count of 117. All layouts in the finer sweep are simulated on the same bank of 600 impacts under correlated background matching, our primary decoder. The initial sweep and the choice of counts for the finer sweep are described in Methods. As expected, the excess risk of the typed layout decreases as more ancillas are protected, crossing the shallow-band reference between 117 and 123 protected ancillas (Fig.~\ref{fig:placement}b; Table~\ref{tab:count-refinement}). Using linear interpolation in log excess risk between simulated counts, we obtain $118.8^{+14.3}_{-1.8}$ protected ancillas at the reference risk. The protected-ancilla saving is $\Delta N=49^{+2}_{-14}$ relative to the shallow band. At the simulated count of 123, the corresponding point-estimate saving is 45. This saving counts protected ancillas, which can be converted to fabrication overhead only after device-specific per-site costs and placement constraints are determined.

\fig{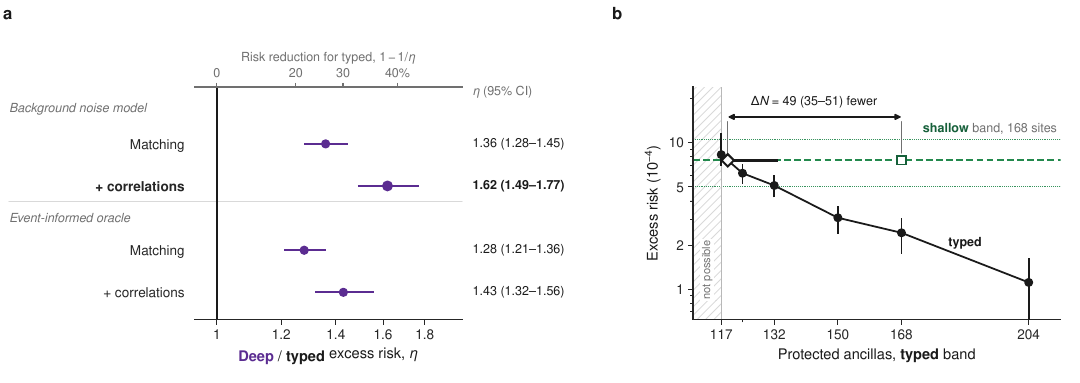}{\textbf{Placement comparisons and protected-ancilla counts at distance 9.} \textbf{a}, Deep-to-typed excess-risk ratios $\eta$ for four decoders, grouped by decoder weights, using 3,200 impacts and 4,096 trajectories per impact, input mode and layout. Ratios above one indicate lower typed risk, and the ratio axis is logarithmic. Values beside each row give $\eta$ and its 95\% interval; the top axis gives the corresponding reduction in typed excess risk, $1-1/\eta$. The 38\% reduction with correlated background matching (bold) has a 95\% interval of 33--44\%. \textbf{b}, Typed excess risk versus protected count with correlated background matching, using 600 impacts and 1,024 trajectories per impact, input mode and layout. The dashed line gives the shallow-band reference at 168 protected sites (open square), and dotted lines give its interval. The diamond marks the log-risk-interpolated count of 118.8 (117.0--133.1) at which typed reaches the reference, with its interval drawn along the reference line; the arrow gives the protected-ancilla saving $\Delta N=49$ (35--51). At the directly simulated count of 123, the point-estimate saving is 45 sites; 28\% of the bootstrap draws reach the shallow reference already at 117, the smallest permitted count. Counts below 117 cannot be realized with the five-layer typed band (hatched). The risk axis is logarithmic. Error bars give nominal 95\% bootstrap intervals.}{fig:placement}

\section{Temporal boundaries and ancilla propagation}
At a temporal boundary, a check may have no neighbouring measurement round against which its report can be compared. Thus, a report error at the last round can produce a single detector defect instead of a pair~\cite{temporalsurgery,fragile}. We probe this mechanism by applying a report-error pulse to a check of the measured type while keeping all other sites at background noise. For the resulting defect, we retain the correction selected by the background matching decoder and define $W$ as the minimum physical-edge log-odds cost of a fault path with the opposite logical parity (Methods). This cost identifies paths that are less suppressed by the noise induced by an impact.

Figure~\ref{fig:paths}a shows how the minimum opposite-parity path cost changes with report-pulse depth during the distance-9 $ZZ$ merge. At depths 1--3, the lowest-cost paths contain five pulse-affected report edges and one, two and three background edges, respectively. From depth 4, adding background edges makes the pulse-affected paths more costly than a quiet path. In contrast, during the $XX$ merge, paths composed only of pulse-affected edges persist through depth 4, and paths combining pulse and background edges extend through depth 7 (Fig.~\ref{fig:paths}b). We attribute the longer reach to the final $Z$ readout of A: its X-check measurements end without a subsequent comparison round, introducing another temporal boundary. The circuit also permits an X fault on an X-check ancilla controlling a CX gate to propagate to data qubits, altering the syndrome used by the decoder.

We next examine how the path cost changes when faults from ancillas are included. For a Z check of A at depth one, we obtain $W=7.8$ at the reference dose with all response channels active, compared with 10.4 for excess report errors only and 22.7 without burst-induced excess errors (Fig.~\ref{fig:paths}c). These changes arise from the local dose, propagation from ancillas and distance from the seam. To test whether the path calculation predicts which checks benefit from protection, we compare 32 selected swaps of protection between checks of the same type over 600 impacts. The paths rank the selected swaps in agreement with their estimated risk ordering, giving Spearman $\rho=0.77$ under background matching.

\fig{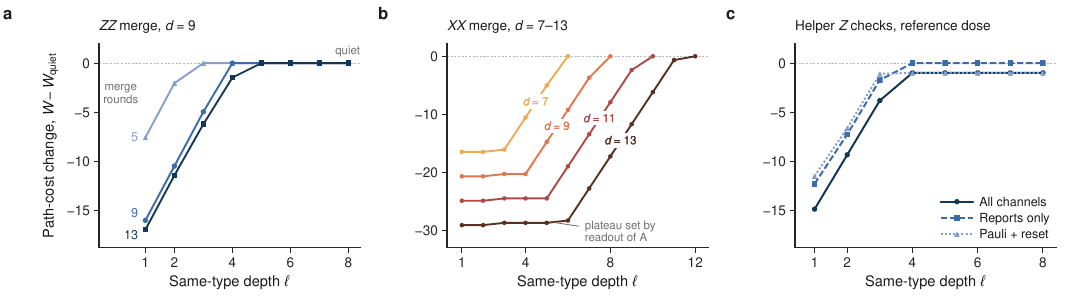}{\textbf{Error-path costs at temporal boundaries.} \textbf{a}, Change in minimum path cost versus same-type check depth for a report pulse during the distance-9 $ZZ$ merge, at three merge durations. \textbf{b}, Corresponding $XX$-merge response for distances 7--13, including the boundary from the final readout of A. Curves are referenced to their quiet-depth plateaus. A pulse gives one check report-flip probability 0.4 during active merge rounds, while other sites remain at background noise. \textbf{c}, Reference-dose response for Z checks of the helper patch, comparing all channels, report excess only, and ancilla Pauli plus reset excess only. Costs use background decoding. Lower cost identifies a less suppressed path; $W$ is an unnormalized path cost.}{fig:paths}

We separate the effects of report errors, Pauli faults of ancillas and reset errors by turning their burst-induced excess contributions on and off in a factorial simulation of 600 impacts, keeping background noise and excess errors on data qubits active in every configuration. Removing excess report errors eliminates the resolved difference in risk between shallow and typed layouts, while removing excess Pauli faults of ancillas reduces that difference by about half (Fig.~\ref{fig:channels}a). We find an interaction between report and ancilla Pauli excess of $1.12\times10^{-4}$, equivalent to half the difference-in-differences averaged over configurations with reset on and off (Fig.~\ref{fig:channels}b); this interaction is greater than zero and increases the shallow-minus-typed risk difference under background matching even though the physical couplings are held constant. The two channels act jointly on logical failure.

\fig{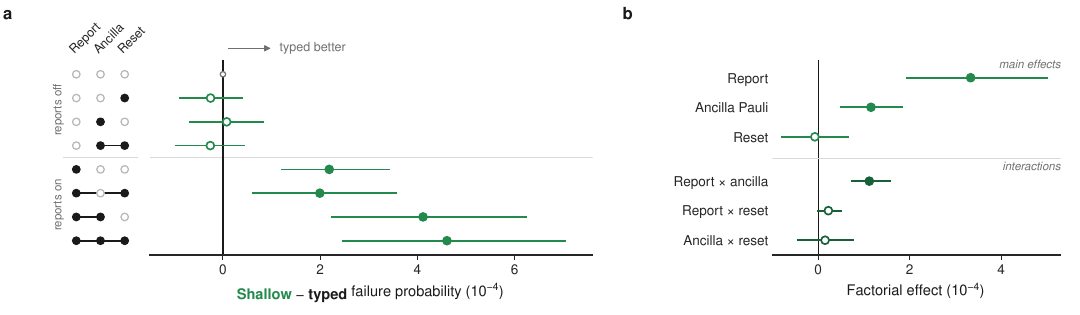}{\textbf{Channel contributions to the shallow-minus-typed risk difference.} \textbf{a}, Input-mean failure-probability difference under background matching for the eight combinations of report, ancilla Pauli and reset burst excess. Filled matrix markers identify active channel families. Data-qubit excess remains active. Positive differences indicate lower risk for typed; the zero point corresponds to identical layouts when all three ancilla channels are disabled. \textbf{b}, Standard factorial main effects and two-family interactions. The main effects of report excess and ancilla Pauli excess are $3.33\times10^{-4}$ and $1.15\times10^{-4}$, respectively. The report--ancilla interaction is $1.12\times10^{-4}$, with an interval of $(0.71\text{--}1.60)\times10^{-4}$; it is half the difference-in-differences averaged over reset-on and reset-off configurations. Whiskers denote 95\% paired event-bootstrap intervals.}{fig:channels}

\section{Decoder information and burst duration}
We next test the typed and deep layouts at distance 13 using 1,200 impacts and two oracle decoders. The ratios of deep to typed excess risk are 0.96 and 1.12, and neither oracle comparison is resolved (Fig.~\ref{fig:robustness}a). In a separate cohort of 2,000 impacts, we also assess three stochastic response models. Typed has lower risk under both background decoders in all three models, while the oracle comparisons are not resolved; the dependence of this ordering on code distance should be the subject of further study. This pattern is consistent with part of the distance-9 placement effect arising from a mismatch between burst errors and background decoding.

In Fig.~\ref{fig:robustness}b,c, we cap burst-induced excess errors after 2, 5, 10 or 20 rounds to test how duration affects protection placement, using background matching and an event-informed oracle without correlations between joint detection events. With a five-round cap, the ratio of shallow to typed excess risk is $0.60^{+0.37}_{-0.32}$ under background matching, indicating that burst duration can reverse the no-cap ordering at matched protection counts. However, the oracle comparison at five rounds and the other finite-cap comparisons are not resolved, except at 20 rounds, where background matching gives lower risk for typed. Finally, the no-cap reference from the same impact bank favors typed under both decoders.

\fig{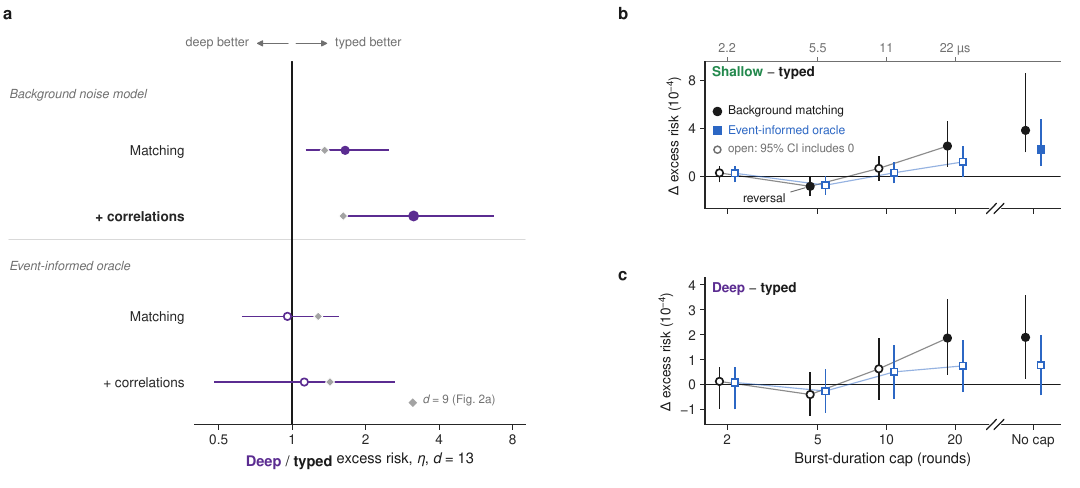}{\textbf{Decoder information and burst-duration comparisons.} \textbf{a}, Deep-to-typed excess-risk ratios at distance 13 using 1,200 impacts, with a logarithmic ratio axis and rows as in Fig.~\ref{fig:placement}a. Grey diamonds repeat the distance-9 ratios of Fig.~\ref{fig:placement}a. The corresponding 95\% intervals are 0.62--1.56 for the oracle ratio of 0.96 and 0.48--2.65 for the correlated oracle ratio of 1.12. Both intervals contain one; open markers denote intervals containing one. \textbf{b},\textbf{c}, Shallow-minus-typed (\textbf{b}) and deep-minus-typed (\textbf{c}) excess-risk differences at distance 9 for burst caps of 2, 5, 10 and 20 rounds. The no-cap reference uses report scale one and the same 600-impact bank, masks and seeds, but is not a fifth finite duration. Positive differences indicate lower typed risk. The five-round shallow comparison is reversed (labeled). The top axis of \textbf{b} converts rounds to burst duration at 1.1~$\mu$s per round, and the axis break separates the no-cap reference. Only background matching and the event-informed oracle are included. The 3,200-impact no-cap cohort in Fig.~\ref{fig:placement}a uses the same noise law but is analyzed separately. In panel b, the background-matching intervals at 2 and 10 rounds contain zero, and the oracle intervals at every finite cap contain zero. Whiskers are nominal 95\% bootstrap intervals; open markers denote intervals containing zero.}{fig:robustness}

The duration of an impact-induced error depends on its physical channel. In earlier superconducting arrays, impact-induced relaxation lasted approximately 25\,ms~\cite{mcewen}, while a gap-engineered array shows excitation bursts ending within about 5\,$\mu$s, longer relaxation bursts lasting tens of microseconds and correlated frequency shifts persisting for approximately 1\,ms~\cite{phase}. Gap engineering changes the strength and duration of relaxation by different factors~\cite{pinckney}. In our transport model, each round lasts 1.1\,$\mu$s, so a cap of five rounds corresponds to 5.5\,$\mu$s and represents a fast response relative to the longer-lived frequency shifts. The stochastic Pauli description is least validated for coherent frequency shifts over longer durations, the regime in which the placement advantage is resolved. A device-calibrated response across these time scales is required to test whether the simulated ordering persists on hardware.

An estimated-burst decoder reduces excess risk conditional on an impact, but false alarms can introduce errors during quiet operation. From its estimated false-alarm rate, we obtain a break-even burst probability of about 4.3\% per operation; at the rare impact rates measured in superconducting arrays~\cite{mcewen,phase}, the point estimates instead predict that false alarms dominate. Uncertainty in the false-alarm estimate remains large, leaving the net benefit unclear.

\section{Outlook}
We have shown in simulation that protecting the check type measured during lattice surgery can reduce the logical risk of high-energy impacts relative to geometric protection at matched budgets. At distance 9, typed protection reduces excess risk by 38\% under correlated background matching, despite matched protected-ancilla counts and X:Z splits. Temporal boundaries in the measurement record make paths sensitive to report errors, while faults on ancillas propagate to data qubits; these mechanisms give rise to an interaction between the two channels in decoded logical risk. At the shallow-band reference risk, the typed layout uses an estimated $\Delta N=49$ fewer protected ancillas (Fig.~\ref{fig:placement}b).

Even so, the protection layouts are evaluated with stochastic response models that have yet to be validated on a device. Physical constraints on gap engineering, coherent phase accumulation and per-site protection costs could change the ordering or usable saving~\cite{gap,pinckney}. The saving estimate applies to the five-layer typed rule, its permitted counts and the reference risk under the selected decoder. Hardware tests should characterize impact-driven errors during the $ZZ$ and $XX$ merges, their duration and the achievable per-site protection contrast, then evaluate placement with background and event-informed decoding across code distances. If the measured-type advantage persists under calibrated device response, protection could be assigned by the circuit rather than by geometry.

\section{Methods}\label{sec:methods}
\subsection{Circuit, noise and protection}
The rotated-surface-code CNOT is implemented with $ZZ$ and $XX$ merges of duration $T_M=d$, separated by $d$ rounds of memory. In input mode Z, the state $|00\rangle$ is prepared and the observables $Z_C$ and $Z_CZ_T$ are tested; in input mode X, $|{+}{+}\rangle$ is prepared and $X_T$ and $X_CX_T$ are tested. A failure is counted when any tested logical output is flipped. The circuit includes 12 rounds before and 12 rounds after the operation, with readout of the helper patch after the $XX$ merge. We implement the sequence directly, outside the standard sequence provided by the simulator.

The background circuit noise follows SI1000 at $p=10^{-3}$~\cite{si1000}. Impact positions are sampled uniformly, energies log-uniformly from 0.5--10\,keV and onset times uniformly through the end of the $XX$ merge. These parameters are mapped to local doses using a finite-chip transport model with a spatial domain extending 1.15\,mm beyond the bounding box of the sites. Gap parameters of 12 and 5\,GHz are used for strong and weak sites, respectively. Within each channel, the alternatives for Pauli errors are categorical, the probability of a report flip depends on the measurement bit and excess reset errors follow the readout model. The stochastic response has not yet been calibrated to a device.

Measured-type bands are protected to a width of $\mathrm{round}(7d/13)$ in layers of checks of the same type. Weak-site assignments displaced by the bands are reassigned in a deterministic geometric order within each check type. Shallow and deep bands protect both check types by distance from the seam, with the deepest permitted band used for deep. Data qubits and sites used only during merges are strongly protected. At each budget, the protected counts and X:Z splits are matched across layouts. The split varies between budgets.

\subsection{Risk and decoding}
We denote the failure probability for layout $S$, impact $e$, input mode $\beta$ and decoder $D$ by $r_D^\beta(S,e)$ and denote the corresponding quiet baseline by $q_D^\beta$. We define the excess risk as
\begin{equation}
E_D(S)=\max_{\beta\in\{Z,X\}}\left[\mathbb E_e r_D^\beta(S,e)-q_D^\beta\right],
\label{eq:excess}
\end{equation}
where the input mode giving the maximum is selected again in each bootstrap draw. The placement ratio is $\eta_D=E_D(\mathrm{deep})/E_D(\mathrm{typed})$, and the protected-ancilla saving $\Delta N$ is the shallow-band count minus the typed count at the reference risk. For factorial tests and local swaps, we use the average over input modes. Detector values from each simulated record are provided to the decoders. The event-informed weights depend on the realized impact and on state-averaged report probabilities, with sampled fault histories excluded. Correlated matching is enabled during construction and application of the decoder~\cite{correlated}. We use Stim 1.16 and PyMatching 2.4~\cite{stim,pymatching}.

\subsection{Sampling and uncertainty}
Within a simulation cohort, the same impacts and paired trajectory seeds are used for all layouts, and each decoder operates on the same trajectories. Trajectories are allocated across dose strata to increase sampling of informative impacts while preserving the weights of the source distribution. For source-pool weight $\widehat\omega_h$, $n_h$ impacts and $K_{ha}$ failures among $N_{ha}$ trajectories, the mode-specific risk is estimated as
\begin{equation}
\widehat R=\sum_h\widehat\omega_h\frac{1}{n_h}\sum_{a=1}^{n_h}\frac{K_{ha}}{N_{ha}}.
\label{eq:estimator}
\end{equation}
We calculate uncertainty intervals using 4,000 bootstrap draws of events within strata, preserving pairing of layouts and decoders, with draws from a Dirichlet distribution for uncertainty in the source-pool weights. For placement and finer-count results, Jeffreys beta draws also account for uncertainty in the quiet binomial baseline; earlier duration intervals use a constant quiet estimate. When a denominator of excess risk approaches zero, we report a difference in place of a ratio. Intervals are nominal and unadjusted unless otherwise specified.

We size the replication of the placement comparison after an earlier, less precise result and report its estimate separately. A control with four seeded events in the current software environment does not reproduce the earlier histograms of individual trajectories. Seeded sampling in Stim can depend on the machine, although the cause of this difference has not been identified. The initial protection-count sweep covers typed budgets of 132, 168 and 204 sites; typed reaches the shallow reference at the lowest tested count under both correlated decoders, giving only a lower bound on the saving. We then run every configuration of the finer count sweep jointly without reusing the older trajectories, and evaluate only correlated background matching in that sweep. For the other three decoders, the initial-sweep estimates remain separate. The measured running time of correlated matching is roughly seven to eight times that of uncorrelated matching in our implementation.

\subsection{Path calculations and count interpolation}
For each physical graph edge $j$, we define cost $\lambda_j=\log[(1-p_j)/p_j]$. A correction is selected using the decoder metric, and physical paths are assigned costs using the event-weighted metric. We minimize the latter over paths from a boundary to a detector defect with logical parity opposite to the selected correction, excluding ties in the decoder metric. The minimum cost $W$ is unnormalized. At the reference dose, we combine the contributions of report, ancilla Pauli and reset errors after Clifford propagation in calculations for a single site. For the selected swaps, the interval for Spearman $\rho=0.77$ is 0.53--0.82.

For the finer count sweep, typed layouts with 117, 123, 132, 150, 168 and 204 protected sites and a shallow reference of 168 sites are simulated using the original bank of 600 impacts. We use event-pool seed 78001 and stratum sizes 300/180/120, with 1,024 trajectories for each event, input mode and layout, giving 8.6 million trajectories in total. The proposed counts of 96 and 114 sites are outside the possible configurations of a five-layer typed band and are excluded. After timing a separate control of four events, and before assessing any risk estimate in the finer sweep, we restrict decoding to correlated background matching. Between adjacent simulated counts, we interpolate linearly in log excess risk and choose the first typed count reaching the shallow reference; neither a monotonic fit nor extrapolation is used. Draws that reach the reference at 117 sites are assigned 117. For each of 4,000 paired bootstrap draws, we resample impacts within strata, draw source weights from a Dirichlet distribution and shared quiet baselines from a Jeffreys distribution, then select the input mode with the larger excess risk. The intervals quantify sampling uncertainty under this procedure. Fabrication overhead requires a separate model.

\section*{Funding}
No external funding was received.

\section*{Data and code availability}
Code, per-event failure counts, and analysis scripts supporting the reported results are available at \href{https://doi.org/10.5281/zenodo.24046121}{Zenodo, DOI: 10.5281/zenodo.24046121}~\cite{zenodo}. The archive README identifies values reproduced from stored analyses; full sampled trajectories are not included.

\end{document}